\documentclass[useAMS,usenatbib]{mn2e}

\usepackage{latexsym}
\usepackage{amssymb}
\usepackage[dvips]{graphicx}
\usepackage{mathrsfs}

\newcommand{\mbh}{$M_{\rmn{BH}}$}
\newcommand{\rblr}{$R_{\rmn{BLR}}$}
\newcommand{\mmbh}{M_{\rmn{BH}}}
\newcommand{\mrblr}{R_{\rmn{BLR}}}
\newcommand{\mlogm}{\log(M_{\rmn{BH}}/\rmn{M}_{\odot})}
\newcommand{\vblr}{$v_{\rmn{BLR}}$}
\newcommand{\mvblr}{v_{\rmn{BLR}}}
\newcommand{\mmsun}{\rmn{M}_{\odot}}
\newcommand{\mbhlb}{$M_{\rmn{BH}} - L_{\rmn{bulge}}$}

\newcommand{\cq}{C\textsc{iv}}

\newcommand{\lambdallambda}{$\lambda L_{\lambda}(1350\rmn{\AA})$}

\title[The BH mass of nearby QSOs: a comparison of the bulge
luminosity and virial methods]{The BH mass of nearby QSOs: a comparison of the bulge
luminosity and virial methods}
\author[M. Labita, A. Treves, R. Falomo and M. Uslenghi]{M. Labita$^{1}$\thanks{E-mail:
marzia.labita@poste.it}, A. Treves$^{1}$, R. Falomo$^{2}$ and M.
Uslenghi$^{3}$\\
$^{1}$Department of Physics and Mathematics, University of Insubria, Via Valleggio 11, I-22100 Como, Italy\\
$^{2}$INAF, Astronomical Observatory of Padova, Vicolo dell'Osservatorio 5, I-35122 Padova, Italy\\
$^{3}$INAF -- IASF, Via Bassini 15, I-20133 Milano, Italy}
\begin{document}

\date{Accepted ... Received ...; in original form ...}

\pagerange{\pageref{firstpage}--\pageref{lastpage}} \pubyear{0000}

\maketitle

\label{firstpage}

\begin{abstract}
We report on the analysis of the photometric and spectroscopic properties of a sample of 29 low redshift ($z<0.6$) QSOs for which both \textit{HST} WFPC2 images and ultraviolet \textit{HST} FOS spectra are available.
For each object we measure the $R$ band absolute magnitude of the host galaxy, the \cq
(1550~\AA) line width and the 1350~\AA{} continuum luminosity.
From these quantities we can estimate the black hole (BH) mass through the $M_{\rmn{BH}}-L_{\rmn{bulge}}$ relation for inactive galaxies, and from the virial method based on the kinematics of the regions emitting the broad lines.
The comparison of the masses derived from the two methods yields information on the geometry of the gas emitting regions bound to the massive BH.
The cumulative distribution of the line widths is consistent with that produced by matter laying in planes with inclinations uniformly distributed between $\sim$10$^{\circ}$ and $\sim$50$^{\circ}$, which corresponds to a geometrical factor $f\sim1.3$. Our results are compared with those of the literature and discussed within the unified model of AGN.
\end{abstract}

\begin{keywords}
galaxies: active -- galaxies: nuclei -- quasars: general.
\end{keywords}

\section{Introduction}
Nowadays there is wide consensus that most if not all nearby early type galaxies host a supermassive black hole (SMBH) at their centers (see for a recent review Ferrarese 2006). However a direct determination of the BH mass (\mbh) is possible only for a limited number of inactive galaxies in the local Universe, from dynamical methods (virial theorem) based on the observations of the kinematics of stars 
orbiting around the SMBH. 
An extension of this method to higher redshift objects is viable only for active galaxies, 
where \mbh{} can be derived from the dynamics of gas gravitationally bound to the SMBH: this is the usual case of regions emitting the broad emission lines (BLR). 

At low redshift correlations between \mbh{} and the global properties of the host galaxies, namely the luminosity ($L_{\rmn{bulge}}$) and the central velocity dispersion of the bulge, have been found. Locally therefore the measure of the properties of the host galaxy is an alternative method for evaluating \mbh.

The aim of this paper is to compare the mass determinations through the virial method and through the global parameters of the hosts of close--by quasars.
For QSOs it is very difficult to measure the velocity dispersion of the stars in the bulge, so one must rely on the \mbhlb{} relationship. 
For nearby ellipticals, there is increasing evidence that 
this relation has the same behavior for both inactive 
and active galaxies 
(see for example Merritt \& Ferrarese 2001; McLure \& Dunlop 2001, 2002; Bettoni et al. 2003), notwithstanding a substantial increase of the dispersion in the latter case, possibly due to the greater difficulty in the determination of all the parameters.
Moreover, AGN hosts and inactive galaxies appear substantially the same objects, as indicated by direct morphological studies (e.g. Dunlop et al.~2003). This is consistent with a picture where  the AGN phenomenon, which is linked to the gas accretion on the central BH, would be a phase characterizing the life of normal galaxies in particular during major mergers (see for example Kauffmann \& Haehnelt 2000).
For these reasons, in this work we start with the assumption that the correlation established for nearby galaxies is also valid for higher redshift QSOs, provided that host galaxy luminosity is corrected for galactic evolution to $z\sim0$. 

A second set of QSO BH mass determinations can be obtained through the virial theorem, provided that an estimate of the velocity \vblr{} and radius \rblr{} of the BLR is available. The BLR velocity can be derived from the FWHM of the broad emission lines if one assumes a value for the geometrical factor $f$, which is an unknown parameter of order unity linked to the geometry of the BLR. The measurement of \rblr{}
may require cumbersome procedures, like reverberation mapping (Ulrich et al.~1984; Blandford \& McKee 1982; Peterson 2001), that uses the continuum source variability to evaluate time delays between the central emission and the BLR gas response; an alternative is to infer the BLR size from the ionizing luminosity with which it is correlated (Kaspi et al.~2000; Vestergaard 2002; Pian, Falomo \& Treves 2005; Kaspi et al.~2005; Vestergaard \& Peterson 2006).
BH mass determinations can be then derived from a measure of the broad line widths and of the ionizing continuum intensity, once the value of the geometrical factor is fixed.

In the next section we focus on the quasar sample choice
and on the data collection and analysis (\ref{secima}--\ref{secspe}). 
In section \ref{secmas} we describe the formulae used for applying the two methods of mass determination (\ref{secpho}--\ref{secvir}) and report on the results (\ref{seccomp}). This analysis constrains the value of the geometrical factor. The issue of the BLR geometry and orientation is further discussed in paragraph \ref{calcolof}. In section \ref{secsum} our results are compared with those of literature and discussed within the unified model of AGN.

Throughout this paper, we adopt a concordant cosmology with $H_0=72$ ~km~s$^{-1}$~Mpc$^{-1}$, $\Omega_m=0.3$ and $\Omega_{\Lambda}=0.7$. We have converted the results of other authors to this cosmology when adopting their relations and using their data.

\section{The sample and data analysis}
\label{secsam}
We selected\footnote{This research has made use of the {\it VizieR Service}, available at \texttt{http://vizier.u-strasbg.fr/viz-bin/VizieR}.}  all QSOs  from the Veron--Cetty \& Veron (2003) catalogue, for which {\it HST} images and spectra are available.
We further require that the images have exposure times longer than 1000~s and that spectra cover the range 1400--1700~\AA{} ($UV$ band).
In addition we selected only objects hosted by ellipticals, so that the bulge component substantially coincides with the whole galaxy.

\begin{figure}
\includegraphics[width=0.45\textwidth]{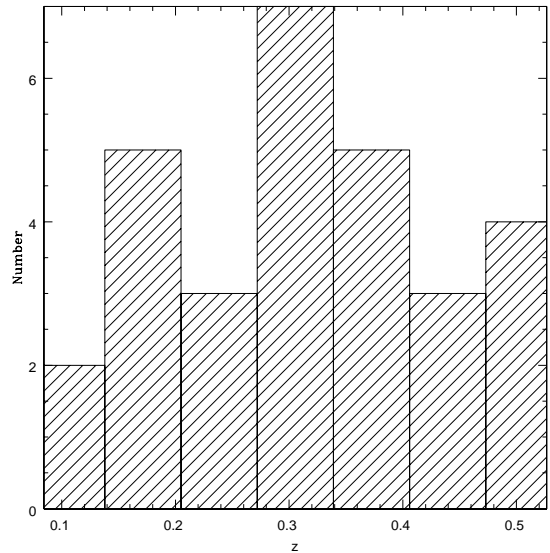}
\caption[]{The redshift distribution of the 29 QSOs with both HST images and $UV$ spectra.}
\label{zdistrib}
\end{figure}

The resulting  sample includes 29 objects at $z<$0.6, 17 of which are radio loud (RLQ) and 12 are radio quiet (RQQ).
The main properties of the objects are listed in Table 1.  The sample is essentially uniformly distributed in redshift  
(see Fig.~\ref{zdistrib}).
\begin{table*}
 \centering
 \begin{minipage}{\textwidth}\label{sample}
  \centering
  \caption{QSOs at $z<0.6$ with {\it HST} images and spectra. These data are taken from the Veron--Cetty \& Veron catalogue. }
\begin{tabular}{@{}ccccccc@{}}
\hline 
\hline
Name		&	Alternative Name&	R.A. (J2000.0)	&	Decl (J2000.0)	&	$z$	&$R_{\textrm{r-o}}$&	$V$	\\
(1) 		&	(2)		& (3) 			& (4) 			& (5) 		& (6) 		   & (8)	\\
\hline
0100+020	&	PHL 0959	&	01 03 12.9	&	+02 21 10.3	&	0.393	&	Q	&	16.39	\\
0133+207	&	3C 047		&	01 36 22.4	&	+20 57 14.7	&	0.425	&	L	&	18.10	\\
0137+012	&	PHL 1093	&	01 39 57.3	&	+01 31 46.2	&	0.272	&	L	&	17.07	\\
%0454--220	&	PKS 0454--22	&	04 56 08.9	&	--21 59 09.0	&	0.533	&	L	&	-	\\
0624+691	&	HS 0624+6907	&	06 30 08.6	&	+69 05 40.0	&	0.370	&	Q	&	14.20	\\
%0637--752	&	PKS 0637-75	&	06 35 46.5	&	--75 16 16.8	&	0.651	&	L	&	15.75	\\
0850+440	&	US 1867		&	08 53 34.2	&	+43 49 01.0	&	0.513	&	Q	&	16.40	\\
0903+169	&	3C 215		&	09 06 31.9	&	+16 46 11.5	&	0.412	&	L	&	18.27	\\
1001+291	&	TON 28		&	10 04 06.1	&	+28 55 19.2	&	0.330	&	Q	&	15.50	\\
1136--135	&	PKS 1136-13	&	11 39 10.7	&	--13 50 43.6	&	0.554	&	L	&	16.17	\\
1150+497	&	4C +49.22	&	11 53 22.3	&	+49 30 21.4	&	0.334	&	L	&	17.10	\\
1202+281	&	PG 1202+281	&	12 04 42.1	&	+27 54 11.0	&	0.165	&	Q	&	15.60	\\
1216+069	&	PG 1216+069	&	12 19 23.1	&	+06 38 26.8	&	0.331	&	Q	&	15.65	\\
1219+755	&	MRK 205		&	12 21 45.9	&	+75 19 06.5	&	0.071	&	Q	&	15.24	\\
1226+023	&	3C 273		&	12 29 06.7	&	+02 03 08.6	&	0.158	&	L	&	12.90	\\
1230+097	&	LBQS 1230+0947	&	12 33 28.8	&	+09 31 04.9	&	0.415	&	Q	&	16.15	\\
1302--102	&	PG 1302--102	&	13 05 33.0	&	--10 33 19.4	&	0.286	&	L	&	15.20	\\
%1305+069	&	3C 281		&	13 07 54.0	&	+06 42 14.3	&	0.599	&	L	&	17.02	\\
1307+085	&	PG 1307+085	&	13 09 47.0	&	+08 19 48.9	&	0.155	&	Q	&	15.10	\\
1309+355	&	TON 1565	&	13 12 16.3	&	+35 14 36.7	&	0.184	&	L	&	15.64	\\
1416--129	&	PG 1416--129	&	14 19 05.7	&	--13 10 56.5	&	0.129	&	Q	&	16.10	\\
1425+267	&	TON 202		&	14 27 33.6	&	+26 32 52.9	&	0.366	&	L	&	15.68	\\
1444+407	&	PG 1444+407	&	14 46 45.9	&	+40 35 05.0	&	0.267	&	Q	&	15.70	\\
1512+370	&	PG 1512+370	&	15 14 39.2	&	+36 50 37.7	&	0.370	&	L	&	16.27	\\
1545+210	&	3C 323.1	&	15 47 43.5	&	+20 52 16.7	&	0.264	&	L	&	16.70	\\
1641+399	&	3C 345		&	16 42 58.8	&	+39 48 37.0	&	0.594	&	L	&	15.96	\\
1704+608	&	3C 351		&	17 04 41.4	&	+60 44 30.5	&	0.371	&	L	&	15.28	\\
1821+643	&	HB89 1821+643	&	18 22 02.8	&	+64 20 05.3	&	0.297	&	Q	&	14.10	\\
2128--123	&	PKS 2128--12	&	21 31 35.3	&	--12 07 04.8	&	0.501	&	L	&	16.11	\\
2135--147	&	PKS 2135--14	&	21 37 45.2	&	--14 32 55.8	&	0.200	&	L	&	15.50	\\
2141+175	&	HB89 2141+175	&	21 43 35.5	&	+17 43 48.0	&	0.213	&	L	&	15.73	\\
2201+315	&	4C +31.63	&	22 03 15.0	&	+31 45 38.3	&	0.295	&	L	&	15.58	\\
\hline
\end{tabular}
\end{minipage}
\end{table*}

\subsection{\textit{HST} WFPC2 images}\label{secima}
For 27 sources in the sample a measurement of the host galaxy magnitude and morphology is available in the literature; while for 2 objects the determination of the parameters of the host galaxy is obtained in this work: the images reduction and analysis is discussed in Appendix A and the relevant data are reported in Table 5.
In order to obtain a homogeneous set of absolute magnitudes of the host galaxies, we collect from literature the values of apparent magnitude (see Table \ref{absolute} for references). Since Floyd et al. (2004) didn't report the values of apparent magnitude for their objects, we derived them from absolute values and corrections they applied.
For all the objects in our sample we apply
color transformations to the $R$ band (following Fukugita, Shimasaku \& Ichikawa 1995), Galactic extinction (following Schlegel, Finkbeiner \& Davis 1998\footnote{This research has made use of the NASA/IPAC Extragalactic Database (NED) which is operated by the Jet Propulsion Laboratory, California Institute of Technology, under contract with the National Aeronautics and Space Administration.}) and $k$--correction (following Poggianti 1997), assuming that the host galaxies are dominated by old stellar population (see Table \ref{absolute}). If we suppose that the host galaxies have a significant fraction of a young ($<1\,$Gyr) population, the color transformation and $k$--correction would produce magnitudes that are systematically smaller by $\sim0.15\,$mag. Anyway, there is not strong evidence that QSO hosts in the nearby Universe ($z\lesssim0.5$) have a significative young stellar component (Nolan et al.~2001).

\begin{table}
\begin{center}
\caption[]{Host galaxies absolute magnitudes of the 29 QSOs. \textsc{Notes} -- (2) 1. Hooper, Impey \& Foltz 1997; 2. Hamilton, Casertano \& Turnshek 2002; 3. Floyd et al.~2004; 4. Pagani, Falomo \& Treves 2003; 5. Dunlop et al.~2003; 6. Boyce et al.~1998; 7. Bahcall et al.~1997; 8. Kirhakos et al.~1999; 9. this work. -- (3) $R$-band mean apparent magnitudes of the host galaxies corrected for Galactic extinction.
-- (4) $k$--correction values following Poggianti (1997). -- (5) Elliptical galaxy passive evolution correction [mag] following Bressan et al.~(1994). -- (6) Host galaxies absolute magnitudes corrected for passive evolution.}
\begin{tabular}{@{}cccccc@{}}
\hline 
\hline
Name        &Authors&$R_{ext}$  & $k(R)$  & $E(R)$ & $M_R$	    \\
(1)	    &(2)    &(3)  &  (4)   & (5) & (6)	    \\
\hline
0100+020    &1, 2   &19.83& 0.47 & -0.36   & -21.90	\\ 
0133+207    &2, 4   &19.23& 0.54 & -0.39   & -22.73	 \\ 
0137+012    &2, 5, 6&17.42& 0.28 & -0.23   & -23.34	 \\ 
%0454--220  &7      &20.74& 0.81 & -0.50   & -42.71	\\ 
0624+691    &3      &17.01& 0.43 & -0.33   & -24.57	\\ 
%0637--752   &9     &19.47& 1.19 & -0.61   & -24.04	\\
0850--144   &9      &19.13& 0.75 & -0.48   & -23.44	\\
0903+169    &2, 4   &19.07& 0.51 & -0.38   & -22.81	\\ 
1001+291    &3	    &17.82& 0.36 & -0.29   & -23.45	\\ 
1136--135   &9      &19.12& 0.87 & -0.52   & -23.73	\\
1150+497    &3      &17.59& 0.37 & -0.30   & -23.69	\\ 
1202+281    &2, 7   &17.14& 0.17 & -0.17   & -22.37	\\ 
1216+069    &2, 6   &18.96& 0.37 & -0.29   & -22.32	\\ 
1219+755    &2      &15.14& 0.06 & -0.08   & -22.57	\\ 
1226+023    &2, 7   &15.43& 0.16 & -0.16   & -23.93	\\ 
1230+097    &3      &17.96& 0.52 & -0.38   & -23.93	\\ 
1302--102   &2, 6, 7&17.42& 0.31 & -0.26   & -23.44	\\ 
%1305+069    &9     &19.72& 1.03 & -0.56   & -23.39	\\
1307+085    &2, 7   &17.50& 0.15 & -0.16   & -21.85	\\ 
1309+355    &2, 4, 7&16.51& 0.19 & -0.18   & -23.27	\\ 
1416--129   &2      &17.42& 0.12 & -0.14   & -21.44	\\ 
1425+267    &2, 8   &18.44& 0.42 & -0.33   & -23.08	\\ 
1444+407    &7      &18.01& 0.29 & -0.24   & -22.67	\\ 
1512+370    &2, 4   &18.45& 0.43 & -0.33   & -23.13	\\ 
1545+210    &2, 7   &17.60& 0.28 & -0.24   & -23.07	\\ 
1641+399    &8	    &19.31& 1.01 & -0.56   & -23.82	\\ 
1704+608    &2, 6   &18.00& 0.43 & -0.33   & -23.58	\\ 
1821+643    &3      &16.50& 0.32 & -0.27   & -24.45	\\ 
2128--123   &6      &19.35& 0.71 & -0.47   & -23.14	\\ 
2135--147   &2, 5, 7&16.94& 0.21 & -0.19   & -23.08	\\ 
2141+175    &2, 5   &17.03& 0.23 & -0.20   & -23.12	\\ 
2201+315    &2      &16.65& 0.32 & -0.26   & -24.30	\\ 
\hline
\end{tabular}
\label{absolute}
\end{center}
%(3) The Galactic Extinction value refers to -- (4) K-correction values, following Poggianti et al.~(1997). -- (6) Absolute magnitude magnitude in $R_{Cousin}$ bandpass. -- (7) $R$ band Galactic evolution values, following Bressan et al.~(1998).
\end{table}

In Table \ref{absolute} we report the $R$--band absolute magnitudes ($M_R$)
corrected for passive evolution (following Bressan, Chiosi \& Fagotto 1994) for the 27 objects discussed above and for the 2 measured by us.
Note that we have considered only QSOs hosted by elliptical galaxies; however for 2 objects (1001+291 and 1444+407)  there is disagreement on the morphology of the host galaxy in literature, and our choice of an elliptical morphology follows respectively the arguments of Floyd et al.~(2004) and Bahcall et al.~(1997).
For 19 objects, more than one estimate of the apparent magnitude was available. In these cases we have considered the mean value. 
The average values of the absolute $R$ magnitudes (not corrected for passive evolution) are:
\begin{displaymath}
\begin{array}{l}
\langle M_R\rangle=-23.47\pm0.81\quad\textrm{QSOs};\\
\langle M_R\rangle=-23.68\pm0.44\quad\textrm{RLQs};\\
\langle M_R\rangle=-23.18\pm1.10\quad\textrm{RQQs}
\end{array}
\end{displaymath}
where the uncertainty is the standard deviation. The gap between the RLQ and RQQ host magnitudes is consistent with the results of other researchers (see for example Hamilton et al.~2002).

\subsection{\textit{HST} FOS spectra}\label{secspe}
\begin{figure*}
\centering
\begin{minipage}{\textwidth}
  \centering
\includegraphics[width=0.8\textwidth]{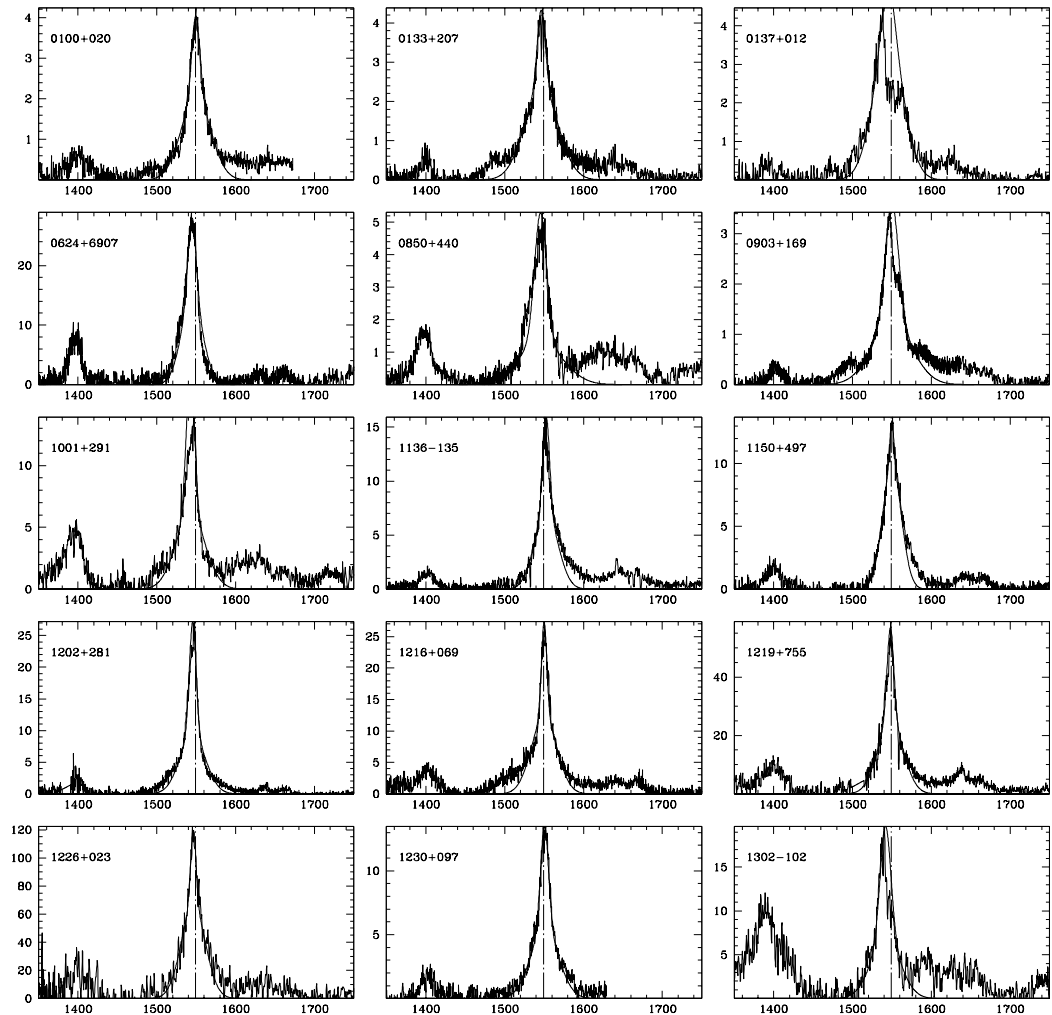} %era 0.7; spettriarticolo
\caption[]{\textit{HST} WFPC2 rest--frame continuum--subtracted spectra: C\textsc{iv} lines. In the $x$-axis, $\lambda$ is expressed in \AA; in the $y$-axis specific flux is expressed in units of $10^{-15}$~erg~cm$^{-2}$~s$^{-1}$~\AA$^{-1}$. %Each pair of vertical lines characterizes a spectral region exploited for the fit. 
The best--fitting function is also plotted for each line.}
\label{spettri}
\end{minipage}
\end{figure*}

\begin{figure*}
\centering
\begin{minipage}{\textwidth}
  \centering
%\caption[]{\textit{HST} WFPC2 rest-frame continuum-subtracted spectra: C\textsc{iv} lines.}
\includegraphics[width=0.8\textwidth]{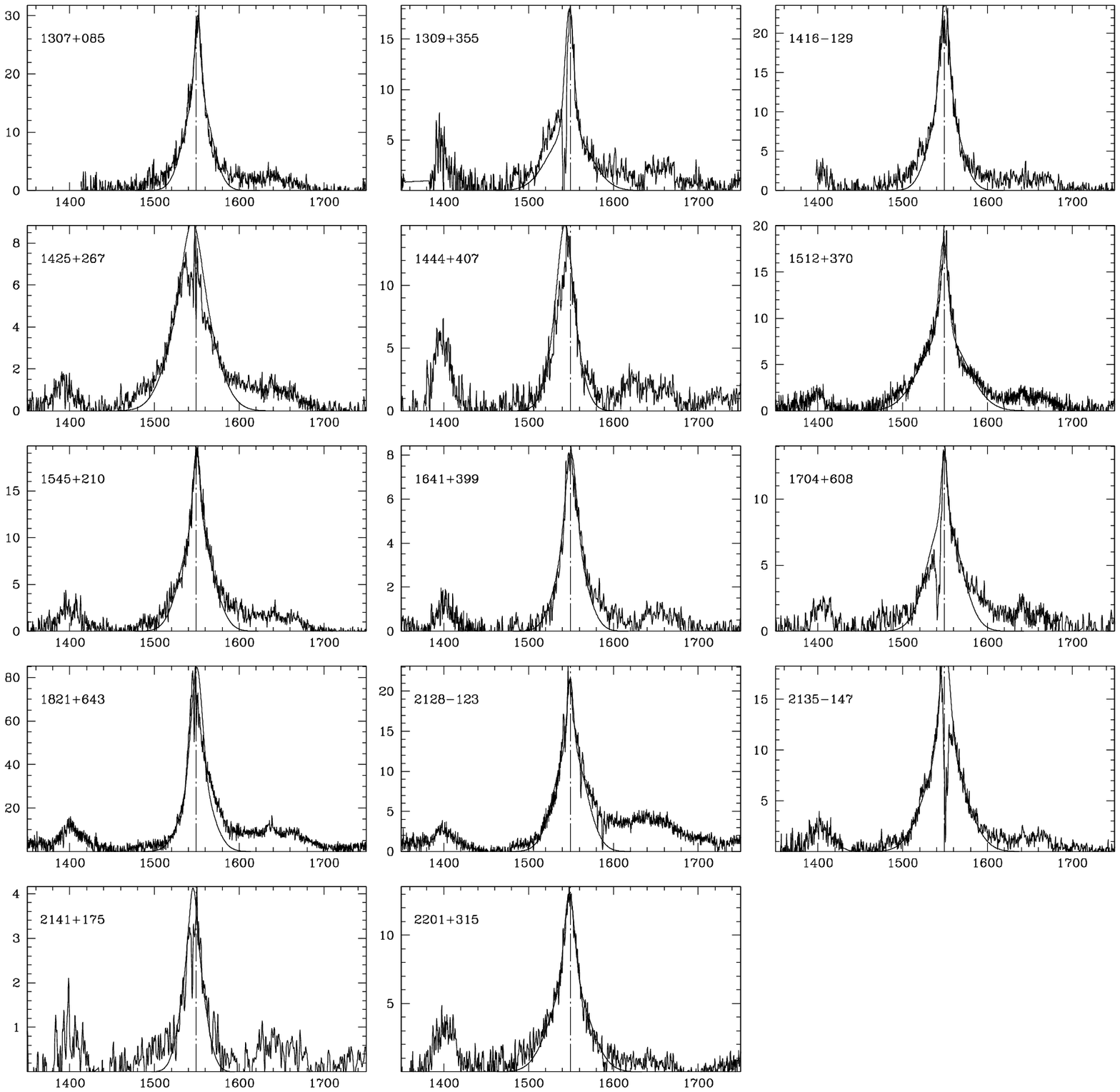} %era 0.7; spettriarticolo
\flushleft{\textbf{Figure \ref{spettri}.} --- continued.}
\end{minipage}
%\textsc{Notes} -- In the $x$-axis, $\lambda$ is expressed in \AA; in the $y$-axis specific flux is expressed in erg~cm$^{-2}$~s$^{-1}$~\AA$^{-1}$. Each pair of vertical lines characterizes a spectral region exploited for the fit. The two Gaussian distributions and their sum are also plotted for each line.
\end{figure*}
In order to have a uniform set of the spectral
parameters, we performed systematically the following procedures to the available data.
When more than one spectrum was available for a single object, different observations were 
combined into a single spectrum.
Then, the continuum was fitted with a power low, removing the regions contaminated by spectral lines and by the Fe\textsc{ii} or He\textsc{ii} emissions. The rest--frame C\textsc{iv} lines of the continuum subtracted spectra are shown in Fig.~\ref{spettri}.

In order to evaluate the virial velocity of the broad line region from $UV$ spectra, we measured C\textsc{iv} line--widths, following i.e. Pian et al.~2005. 
The line widths have been measured by fitting C\textsc{iv} emission lines, excluding the line regions contaminated by absorption features or by the He\textsc{ii}(1640\AA) emission. The lines have been fitted with the sum of two Gaussians with the same central wavelength, and then we derived the FWHM of the sum function. This is just a mathematical tool, and it is not related to a physical decomposition into components of different origin. We tested that, inserting a narrow ($<1500\,$km/s) C\textsc{iv} component, the results remain substantially unchanged, consistently with the numerous evidences that in the $UV$ spectra of QSOs the narrow line components are weak and rare (about 2 per cent occurrence rate, see Wills et al.~1993; Baskin \& Laor 2005). Fig.~\ref{spettri} shows the adopted fits.

Since in the literature virial BH mass estimates are often based on H$\beta$ measurements, it is interesting to compare our FWHMs of \cq{}  with those of H$\beta$. We found H$\beta$ measurements for 25 objects of our sample in the dataset presented by Marziani et al.~2003: the average FWHM of H$\beta$ ($5300\pm2200\,$km$\,$s$^{-1}$) is roughly consistent with that derived from our \cq{} measurements ($4200\pm1100\,$km$\,$s$^{-1}$). %within the 25 per cent.
Note the larger dispersion of the H$\beta$ measurements that is likely due to the non homogeneity of the observations from which these data were derived. 
The consistency between \cq{} and H$\beta$ FWHMs is also supported by comparisons based on larger samples of objects (see Vestergaard et al.~2006 and references therein).

To evaluate the size of BLR through Eq.~\ref{pian}, we need to provide a measure of the ionizing luminosity at 1350~\AA. To this aim, when a direct measure was not possible (3 cases), the value of the continuum flux ($F_{\lambda}$) at the requested wavelength was extrapolated from the best--fitting power law. Continuum intensities were corrected for Galactic extinction, consistently with the previous section, and luminosity at 1350~\AA{} was calculated for each object.

The relevant quantities are reported in Table \ref{specdata}. 
The uncertainty of the continuum level depends on the S/N ratio of the observations and on the intrinsic variability of the sources, giving an error of $\sim10$ per cent on $F_{\lambda}$. On the other hand, the uncertainty on the measured FWHM is dominated by the possible presence of absorption features and contaminating weak He\textsc{ii} emission nearby the C\textsc{iv} line. According to the shape of the C\textsc{iv} line, the errors on the FWHM range from $4$ to $12$ per cent, which illustrates the good quality of the data.
\begin{table}
\begin{center}
\caption[]{Rest--frame spectroscopic data of the 29 analyzed objects. \textsc{Notes} -- (2) Rest--frame \cq{} FWHM values -- (3) Rest--frame flux values -- 
(4) Galactic Extinction values following Schlegel et al.~1998 -- (5) Luminosity at 1350~\AA% -- (6) Error category of the FWHM measurement (see text)
.}
\begin{tabular}{@{}ccccc@{}}
\hline
\hline 
Name	  &  FWHM                  &  $F_{\lambda}$(1350\AA)  & $A_{\lambda}$ &  log$(\lambda L_{\lambda})$\\
\phantom{.}	  & $\left[10^{3}\frac{\rmn{km}}{\rmn{s}}\right]$& $\left[10^{-15}\frac{\rmn{erg}}{\rmn{cm}^{2}\,\rmn{s}\,\rmn{\AA}}\right]$& $[$mag$]$&$\left[\log\frac{\rmn{erg}}{\rmn{s}}\right]$\\
(1)	  &  (2)                   &  (3)		      & (4)	      &  (5)			   \\
\hline
0100+020  &  3.97  &  \phantom{11}1.69  &  0.17	    &	45.15			 \\
0133+207  &  4.48  &  \phantom{11}1.16  &  0.52	    &	45.21			 \\
0137+012  &  6.78  &  \phantom{11}0.81  &  0.23	    &	44.49			 \\
%0454--220&  4.61  &  6.71  &  0.42	    &	46.16			 \\
0624+691  &  3.58  & \phantom{1}16.72  &  0.80	    &	46.35			 \\  
%0637--752&  2.05  &  4.27  &  0.97	    &	46.40			 \\  
0850+440  &  4.22  &  \phantom{11}3.00  &  0.32	    &	45.73			 \\  
0903+169  &  4.99  &  \phantom{11}0.69  &  0.34	    &	44.88			 \\  
1001+291  &  3.07  & \phantom{1}11.33  &  0.18	    &	45.81			 \\  
1136--135 &  2.94  &  \phantom{11}4.19  &  0.40	    &	45.99			 \\  
1150+497  &  3.97  &  \phantom{11}3.20  &  0.17	    &	45.26			 \\  
1202+281  &  2.56  &  \phantom{11}2.37  &  0.17	    &	44.45			 \\  
1216+069  &  2.56  & \phantom{1}11.21  &  0.18	    &	45.81			 \\  
1219+755  &  2.94  & \phantom{1}19.71  &  0.37	    &	44.74			 \\  
1226+023  &  3.20  &108.12  &  0.17	    &	46.05			 \\  
1230+097  &  3.33  &  \phantom{11}5.57  &  0.17	    &	45.72			 \\  
1302--102 &  3.97  & \phantom{1}31.22  &  0.34	    &	46.16			 \\  
%1305+069  &  3.97  &  \phantom{11}2.70  &  0.40	    &	45.87			 \\  
1307+085  &  3.33  &  \phantom{11}9.20  &  0.28	    &	45.03			 \\  
1309+355  &  3.20  & \phantom{1}12.13  &  0.10	    &	45.24			 \\  
1416--129 &  3.84  &  \phantom{11}3.16  &  0.79	    &	44.57			 \\  
1425+267  &  8.06  &  \phantom{11}4.38  &  0.15	    &	45.48			 \\  
1444+407  &  5.12  & \phantom{1}15.91  &  0.11	    &	45.71			 \\  
1512+370  &  3.97  &  \phantom{11}8.17  &  0.18	    &	45.79			 \\  
1545+210  &  4.22  &  \phantom{11}5.66  &  0.34	    &	45.35			 \\  
1641+399  &  4.35  &  \phantom{11}3.93  &  0.14	    &	45.93			 \\  
1704+608  &  5.12  & \phantom{1}10.92  &  0.19	    &	45.91			 \\  
1821+643  &  3.97  & \phantom{1}37.10  &  0.34	    &	46.27			 \\  
2128--123 &  4.73  &  \phantom{11}7.59  &  0.58	    &	46.22			 \\  
2135--147 &  4.86  &  \phantom{11}5.22  &  0.42	    &	45.09			 \\  
2141+175  &  4.22  &  \phantom{11}4.54  &  0.90	    &	45.27			 \\  
2201+315  &  4.73  & \phantom{1}14.90  &  0.99	    &	46.13			 \\  
\hline
\end{tabular}
%\caption[Determinazione delle masse per via spettroscopica.]{\begin{footnotesize}Determinazione delle masse per via spettroscopica.Gli ultimi 5 oggetti sono quelli per i quali non sono ancora noti i parametri della galassia.\end{footnotesize}}
\label{specdata} 
\end{center}
\end{table}

\section{BH mass determination}\label{secmas}
\subsection{BH mass from the \mbhlb}\label{pho}\label{secpho}
In nearby inactive galaxies, the correlation between \mbh{} and the luminosity of the bulge component of the host has been deeply investigated (e.g. Magorrian et al.~1998; Kormendy \& Gebhardt~2001; Bettoni et al~2003).
Here we use the relationship obtained by Bettoni et al.~(2003),
 who adopted our same definition and corrections for the magnitudes.  
The relation corrected for the chosen
cosmology is:
\begin{equation}\label{bett}
\mlogm=-0.50M_R-2.60
\end{equation}
where $M_R$ is the absolute magnitude of the bulge component of the host galaxy. The $rms$ of this fit is 0.39.  This relation is close to that obtained by McLure \& Dunlop (2002) for AGN host galaxies.

Thus, from the absolute magnitude of the host galaxies, corrected for passive evolution (see Table \ref{absolute}), we can deduce the BH masses using Eq.~\ref{bett}. These masses are reported in column (2) of Table \ref{mass}. The mean values are: 
\begin{displaymath}
\begin{array}{l}
\langle\, \log(\mmbh/\mmsun)\,\rangle =8.99\pm0.38\qquad\textrm{QSOs};\\
\langle\, \log(\mmbh/\mmsun)\,\rangle =9.08\pm0.21\qquad\textrm{RLQs};\\
\langle\, \log(\mmbh/\mmsun)\,\rangle =8.86\pm0.52\qquad\textrm{RQQs}
\end{array}
\end{displaymath}
where the uncertainty is the standard deviation.

\subsection{BH mass from virial method}\label{vir}\label{secvir}
Recent studies (Gaskell 1998; Peterson \& Wandel 2000; Gebhardt et al.~2000; Ferrarese et al.~2001) suggest that the black hole gravity dominates radiation pressure effects on the BLR. Assuming this and supposing that the system is in equilibrium, it is possible to use the virial theorem to evaluate the central BH mass:
\begin{equation}\label{viriale}
\mmbh=G^{-1}\,\mrblr\,(\mvblr)^{2}
\end{equation}
where \rblr{} is the radius of the broad line region and \vblr{} is the gas velocity at \rblr. 

Under the hypothesis that the spectral line--width is caused by Doppler effect, the BLR velocity can be expressed as:
\begin{equation}\label{velocita}
v_{\rmn{BLR}}=f\cdot \rmn{FWHM}
\end{equation}
where FWHM refers to the broad emission lines produced in the BLR and is expressed in velocity units, and
$f$ is a factor related to the BLR geometry. For an isotropic field of velocities, $f=\sqrt{3}/2$. The issue of the value of $f$ is discussed later on.

A relation which links \rblr{} to the ionizing luminosity at 5100~\AA{}  was 
calibrated on reverberation measurements (Wandel, Peterson \& Malkan 1999, Kaspi et al.~2000). 
Recently, this phenomenological formula has been extended outside the optical spectral range, giving expressions to derive \rblr{} from the continuum luminosity in the $UV$ and $X$--band (Vestergaard 2002; Pian et al.~2005; Kaspi et al.~2005; Vestergaard \& Peterson 2006).
To infer an estimate of the BLR radius, we adopt  the \rblr--$L_{\lambda}(1350\rmn{\AA})$ 
relation given in Pian et al.~2005 that was calibrated directly on reverberation measures of \rblr:
\begin{equation}\label{raggio}
\mrblr=(22.4\pm0.8)\left[\frac{\lambda L_{\lambda}(1350\rmn{\AA})}{10^{44}\textrm{ erg s}^{-1}
}\right]^{0.61\pm0.02}\rmn{lt-days,}
\end{equation}
which is consistent with that proposed by Kaspi et al.~(2005), based on the study of 32 reverberation mapped AGN.

Using Eq.~\ref{velocita} and \ref{raggio} in the virial theorem (Eq.~\ref{viriale}), one can derive the BH 
mass:
\begin{equation}\label{pian}
\mmbh\!=\!f^{2}\cdot4.4\cdot10^{6}\!\!\left[\frac{\lambda L_{\lambda}(1350\rmn{\AA})}{10^{44}\textrm{ erg s}^{-1}}
\right]^{0.61}\!\!\!\left(\!\!\frac{\rmn{FWHM}
}{1000\,\textrm{km s}^{-1}}\!\!\right)^{2}\!\!\mmsun
\end{equation}
Note that \mbh{} depends strongly on the value of the geometrical factor $f$.

We used the values of \lambdallambda{} and FWHM reported in Table \ref{specdata}. The corresponding mass values are calculated with Eq.~\ref{pian} and are given in Table \ref{mass}, under the assumption of isotropy of the BLR geometry ($f=\sqrt{3}/2$).
\begin{table}
\begin{center}
\caption[]{BH mass determinations. \textsc{Notes} -- (2) \mbh{} from the bulge luminosity (Eq. \ref{bett}). -- (3) \mbh{} through the virial method (Eq. \ref{pian}) adopting $f=\sqrt{3}/2$. -- (3) \mbh{} through the virial method (Eq. \ref{pian}) adopting $f=1.3$.}
\begin{tabular}{@{}cccc@{}}
\hline
\hline
Name           &$\!\log(\mmbh/\mmsun)\!$&$\!\log(\mmbh/\mmsun)\!$&$\!\log(\mmbh/\mmsun)\!$\\  %$\log\frac{\mmbh[\rmn{vir}]}{f^{2}}$
                &from $L_{\rmn{bulge}}$	  &$f=\sqrt{3}/2$&$f=1.3$\\
(1)		&(2)	  &(3)      &(4)	   \\
\hline
0100+020	& 8.35  &  8.41  &  8.77 \\
0133+207	& 8.76  &  8.55  &  8.91 \\
0137+012	& 9.07  &  8.47  &  8.83 \\
%454--220	&  	&        &
0624+691	& 9.69  &  9.05  &  9.41 \\
%0637--752	& 9.42  &  8.60  &  8.96 \\
0850+440	& 9.12  &  8.82  &  9.18 \\
0903+169	& 8.80  &  8.44  &  8.80 \\
1001+291	& 9.12  &  8.59  &  8.95 \\
1136--135	& 9.26  &  8.66  &  9.02 \\
1150+497	& 9.25  &  8.48  &  8.84 \\
1202+281	& 8.58  &  7.60  &  7.96 \\
1216+069	& 8.56  &  8.43  &  8.79 \\
1219+755	& 8.68  &  7.90  &  8.26 \\
1226+023	& 9.37  &  8.77  &  9.13 \\
1230+097	& 9.36  &  8.60  &  8.96 \\
1302--102	& 9.12  &  9.03  &  9.39 \\
%1305+069	& 9.10  &  8.85  &  9.21 \\
1307+085	& 8.32  &  8.18  &  8.54 \\
1309+355	& 9.03  &  8.28  &  8.64 \\
1416--129	& 8.12  &  8.03  &  8.39 \\
1425+267	& 8.94  &  9.23  &  9.59 \\
1444+407	& 8.74  &  8.97  &  9.33 \\
1512+370	& 8.97  &  8.80  &  9.16 \\
1545+210	& 8.94  &  8.59  &  8.95 \\
1641+399	& 9.31  &  8.97  &  9.33 \\
1704+608	& 9.19  &  9.09  &  9.45 \\
1821+643	& 9.63  &  9.09  &  9.45 \\
2128--123	& 8.97  &  9.22  &  9.58 \\
2135--147	& 8.94  &  8.55  &  8.91 \\
2141+175	& 8.96  &  8.54  &  8.90 \\
2201+315	& 9.55  &  9.16  &  9.52 \\
\hline													
\end{tabular}
%\caption[Determinazione delle masse per via spettroscopica.]{\begin{footnotesize}Determinazione delle masse per via spettroscopica.Gli ultimi 5 oggetti sono quelli per i quali non sono ancora noti i parametri della galassia.\end{footnotesize}}
\label{mass} 
\end{center}
%\textsc{Notes} -- CAMBIARE!!!!   (3) \mbh(I) refers to \logm, determined through the \mbh$L_{bulge}$ relation -- (4) \mbh(IIa) refers to \logm, determined through the virial method, assuming $f=\sqrt{3}/2$ -- (5) \mbh(IIb) refers to \logm, determined through the virial method, assuming $f=3/2$ -- (6) Difference between values in col.~(3) and col.~(4) -- (7) Difference between values in col.~(3) and col.~(5) -- In the last line are reported the mean values of col.~(6) and col.~(7).
\end{table}

\subsection{Comparison of BH masses}\label{seccomp}
\begin{figure}
\includegraphics[width=0.45\textwidth]{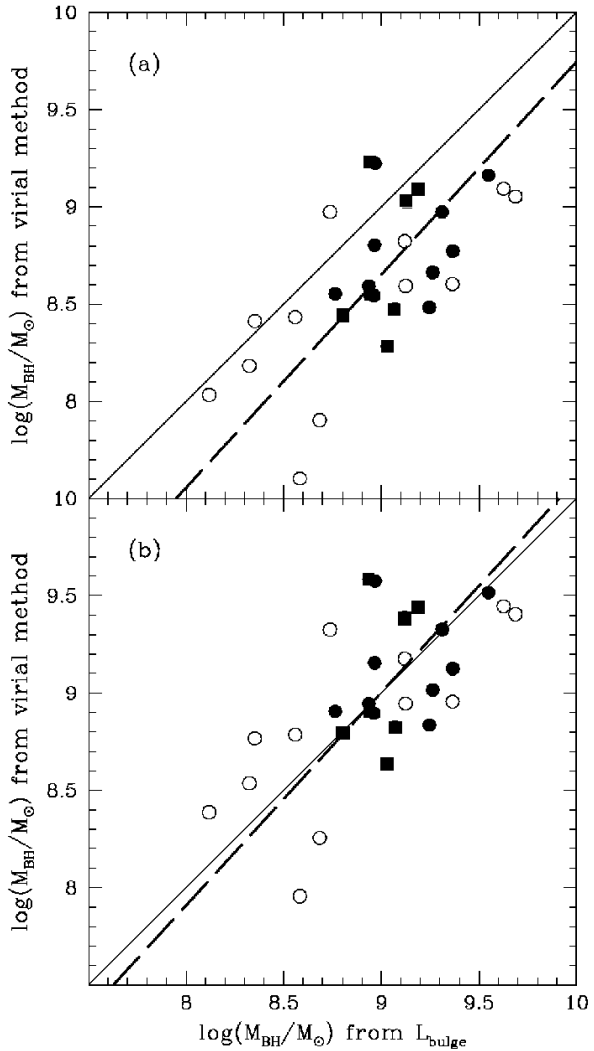}%{mm_k1}%{massemasseok}
\caption[]{Virial BH mass determinations, adopting $f=\sqrt{3}/2$ in panel (a) and $f=1.3$ in panel (b), are plotted against mass values from the \mbhlb{} relation. Filled (open) symbols indicate radio loud (quiet) objects.  Squares highlight the objects with strong absorption features contaminating the \cq{}  line. The solid line is the bisector; the thick dashed line is the best fit to the data.}
\label{massemasseLQ}
\end{figure}

In Fig.~\ref{massemasseLQ} we compare the BH masses obtained with the photometric and virial methods. As clearly shown in panel (a), the simple case of isotropic distribution of the emitting region yields inconsistent results: a systematic difference between the mass determinations obtained with the two methods is apparent  (\mbox{$\,\langle\, \Delta\log(\mmbh/\mmsun)\,\rangle =0.36$}). 

Nevertheless, a linear relation fits well the data point and the slope of the best fit is consistent with unity ($a=1.1\pm0.2$). This suggests that a way of matching the two BH mass estimates is to consider the geometrical factor $f$ as a free parameter, and to search for the value which minimizes the sum of the differences between the masses in logarithmic units from the two methods. This yields $f=1.3\pm0.1$. Assuming this value, we report in Table \ref{mass} the corresponding virial BH masses.

In panel (b) of Fig.~\ref{massemasseLQ} we show the comparison of the BH masses obtained with the photometric and virial methods, adopting $f=1.3$. 

\subsection{Constraining the BLR geometry}\label{calcolof}
The results of the previous paragraph strongly suggest that the simplifying hypothesis of complete isotropy of the BLR, which leads to $f=\sqrt{3}/2$, is to be revised.
Therefore we consider a scenario where there is a component of the BLR velocity field that lies on the plane of the accretion disk (see also McLure \& Dunlop 2002; Jarvis \& McLure 2006), in order to explore the connection between the BLR geometry and the value of the geometrical factor.

Within the unification model, we suppose that AGN are randomly oriented in  space and that they are observed with the characteristics of QSO if they lie with an inclination angle $\theta$ in a range $\theta_{\rmn{MIN}}-\theta_{\rmn{MAX}}$. The probability distribution to observe an object in this range is:
\begin{equation}\label{prob}
P_{\theta}(\theta)=\frac{\sin\theta}{\cos\theta_{\rmn{MIN}}-\cos\theta_{\rmn{MAX}}}
\end{equation}
and $P_{\theta}(\theta)=0$ outside. 

The observed FWHM of the broad emission lines is given by: 
\begin{equation}\label{FW}
\rmn{FWHM}=2\left(  c_1\frac{1}{\sqrt{2}}\sin\theta + c_2\frac{1}{\sqrt{3}}  \right)v_{\rmn{BLR}}
\end{equation}
where $\sin\theta$ accounts for the projection of $v_{\rmn{BLR}}$ on the line of sight and the coefficients $\sqrt{2}$ and $\sqrt{3}$ refer to an isotropic distribution on the plane of the accretion disk and in the 3D space respectively. The parameters $c_1$ and $c_2$ represent the relative fraction of FWHM due to the two velocity components.
Note that, according to Eq.~\ref{velocita}, the mean value of $f$ can be immediately evaluated from Eq.~\ref{FW}:
\begin{equation}\label{fm}
\langle f\rangle =\frac{1}{2\left( \frac{c_1}{\sqrt{2}}\langle \sin\theta\rangle +\frac{c_2}{\sqrt{3}} \right)}
\end{equation}
where $\langle \sin\theta\rangle $ is averaged over the probability distribution of the inclination angle (Eq.~\ref{prob}).
 With these premises, one can infer the expected cumulative FWHM distribution through Eq.~(\ref{prob}) and (\ref{FW}).
%:
%\begin{equation}
%\label{distrib}
%P\,(<\!\rmn{FWHM})=\left\{
%\begin{array}{l}
%0\quad\quad\quad\quad\quad\quad\textrm{if FWHM}<a\\
%\vspace{0.03cm}\\
%\big[(\cos\theta_{\rmn{MIN}}-\cos\theta_{\rmn{MAX}})k_1\big]^{-1}\cdot\\
%\cdot\left[\sqrt{k_1^{2}-(a-k_2)^{2}}+\right.\\
%\left.-\sqrt{k_1^{2}-(\rmn{FWHM}-k_2)^{2}}   \right]\\
%\vspace{0.02cm}\\
%\phantom{1}\quad\quad\quad\quad\quad\quad\textrm{if }a<\rmn{FWHM}<b\\
%\vspace{0.03cm}\\
%1\quad\quad\quad\quad\quad\quad\textrm{if FWHM}>b
%\end{array}\right.
%\end{equation}
%where
%\begin{displaymath}
%\begin{array}{l}
%k_1=2\,c_1\,\langle v_{\rmn{BLR}}\rangle /\sqrt{2}\\
%k_2=2\,c_2\,\langle v_{\rmn{BLR}}\rangle /\sqrt{3}\\
%a=k_1\sin\theta_{\rmn{MIN}}+k_2\\
%b=k_1\sin\theta_{\rmn{MAX}}+k_2\quad\quad.
%\end{array}
%\end{displaymath}

\begin{figure}
\includegraphics[width=0.45\textwidth]{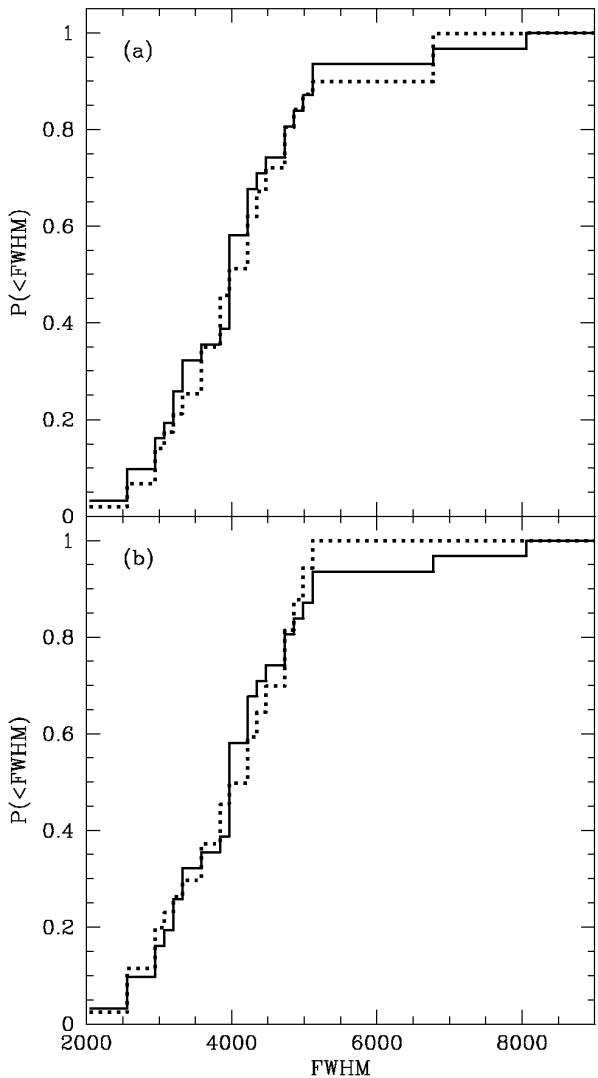}
\caption[]{The cumulative distribution of the observed \cq{} FWHM (solid line) is compared in panel (a) with the Gaussian FWHM distribution which is expected under the assumption of isotropic BLR geometry (dotted line). In panel (b), the same distribution of observed FWHM (solid line) is compared with the expectations in the case of disc--like BLR geometry (dotted line), provided that inclination angles are distributed between $12^{\circ}$ and $48^{\circ}$ (see text for details).}
\label{FWtot}
\end{figure}

In Fig.~\ref{FWtot} we compare the distribution of the observed FWHM of the \cq{}  lines with the expected distribution% %(Eq.~\ref{distrib})
, in the cases of the two extreme assumption of complete isotropy ($c_1=0$) and disc--like BLR geometry ($c_2=0$).

%\paragraph*{Case I -- Isotropic BLR geometry.}
%If
 In the isotropic case 
we assume that the \vblr{} has an intrinsic Gaussian distribution with dispersion $\sigma$. The observed data are consistent with the expected distribution if  
$\frac{\sigma}{\langle \rmn{FWHM}\rangle }=23.4$ per cent, giving a Kolmogorov--Smirnov compatibility test with P(KS)$\sim99.5$ per cent (see panel (a) of Fig.~\ref{FWtot}).
However, this isotropic scenario implies $f\sim\sqrt{3}/2$ (see Eq.~\ref{fm} with $c_1=0$), and therefore it is inconsistent with the comparison of the BH masses through the two methods (paragraph \ref{secvir}).

\begin{figure}
\includegraphics[width=0.45\textwidth]{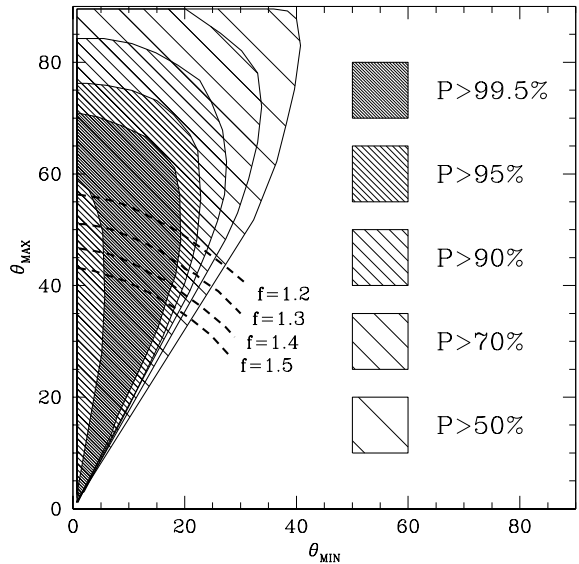}
\caption[]{Kolmogorov--Smirnov probability values for the compatibility between the cumulative distribution of the observed \cq{} FWHM in our sample and the expected distribution constructed under the hypothesis of disc--like BLR geometry, in function of the minimum and maximum inclination angles. The darkest area corresponds to the highest compatibility, with $P$(KS)$>99.5$ per cent. The dashed lines indicate the locus of angles pairs which give the same geometrical factor (see text for discussion).}
\label{contours_def}
\end{figure}

%\paragraph*{Case II -- Disc--like BLR geometry.}
%In this case
 In the case of disc--like geometry% 
, we suppose that the FWHM dispersion is dominated by orientation (i.e. the intrinsic \vblr{} dispersion is negligible). 
Fig.~\ref{contours_def} shows the Kolmogorov--Smirnov probability as a function of $\theta_{\rmn{MIN}}$ and $\theta_{\rmn{MAX}}$. 
Superimposed, we also plot the lines corresponding to fixed values of $f$. It is noticeable that taking $\theta_{\rmn{MIN}}\sim5^{\circ}-20^{\circ}$ and $\theta_{\rmn{MAX}}\sim45^{\circ}-50^{\circ}$ one has P(KS)$>99.5$ per cent (see also panel (b) of Fig.~\ref{FWtot}) and full consistency with the geometrical factor value ($f\sim1.3$) deduced in the previous paragraph.


\section{Summary and conclusions}\label{secsum}
We have evaluated the BH masses of a sample of 29 low redshift quasars using both a photometric and a virial method. 
This comparison suggests the value of the geometrical factor, $f=1.3$.
We also found that the cumulative FWHM distribution of \cq(1550~\AA) is consistent with both a disc--like BLR geometry and an isotropic velocity field, provided that in the latter case there is some distribution around the mean value of the velocity. However, the virial BH masses coincide on average with the photometric ones only in the disc picture, which strongly supports the hypothesis that the broad line region velocity field lies on the plane of the accretion disc. In this scenario, one can also put some constraints on the inclination angles of QSOs, obtaining that they are uniformly oriented between $5^{\circ}-20^{\circ}$ and $45^{\circ}-50^{\circ}$. This result is in good agreement with the unified model of AGN (see Barthel 1989; Antonucci \& Miller 1985; Antonucci 1993; Urry \& Padovani 1995). 

One possible caveat on this result is that we used \cq{} FWHM measurements instead of H$\beta$ ones. If we adopted the H$\beta$ FWHMs (see Section \ref{secspe}), we would obtain on average BH masses $0.2\,$dex larger. This would imply a value of the geometrical factor $f\sim1$, which is still inconsistent with the isotropic picture.

The quasar BH masses range over two decades ($10^{8}\,-\,10^{9.7}\,\mmsun$).
Both the photometric and the virial approaches indicate that RLQ masses exceed $10^{8.6}\mmsun$, while RQQs are distributed in the whole mass range. RLQs have on average BH masses greater than RQQs; the difference $\delta$ between the mean values of $\log\mmbh$ of the RLQ and RQQ subsamples is: 
\begin{displaymath}
\delta\equiv\langle\, \log\mmbh(\rmn{RLQ})\,\rangle -\langle\, \log\mmbh(\rmn{RQQ})\,\rangle =0.26.
\end{displaymath}
This is in good agreement with the results obtained by Dunlop et al.~(2003) and Falomo et al.~(2004), who found respectively $\delta=0.34$ for 23 QSOs at $\langle z\rangle =0.2$ and $\delta=0.33$ for 16 QSOs at higher redshift ($\langle z\rangle =1.5$).
%; however this contrasts with the results of Floyd et al.~(2004), who obtained that RLQs are on average less massive than RQQs, with $\delta=-0.26$ for a sample of 14 QSOs at $\langle z\rangle =0.4$. 

The general agreement between the photometric and the spectroscopic methods may be considered as an indirect indication that the \mbhlb{} correlation obtained locally can be extrapolated at higher redshift ($z\sim0.5$). 

Our results on the BLR geometry agree well with that of the seminal paper of McLure \& Dunlop (2001), who proposed that $f=1.5$ and obtained that a sample of 30 QSOs and 15 Seyfert I galaxies is well described by a disc--like BLR geometry, provided that it is composed of two populations with inclination angles uniformly distributed within $0^{\circ}<\theta<20^{\circ}$ and $0^{\circ}<\theta<46^{\circ}$ respectively. Our set of quasars is comparable to theirs, but the spectroscopy is of much better quality. In fact, the scatter in terms of black hole mass around the best--fitting relation of Fig.~\ref{massemasseLQ} ($\sigma=0.30$) is appreciably smaller than that obtained by McLure \& Dunlop ($\sigma=0.59$). Our results are in good agreement also with the findings by Onken et al. (2004), who compared \mbh{} derived by the virial method with those from the stellar velocity dispersion for a sample of 16 low redshift AGN, obtaining that the geometrical factor, following our definition, is $f\sim1.2$.

In conclusion, we believe that there is strong evidence that, regardless of the method used to derive BH masses, the geometrical factor results in the range $1.0<f<1.3$.

\begin{footnotesize}

\end{footnotesize}
\section*{Appendix A: HST imaging of 2 QSO}
\begin{table*}
 \centering
 \begin{minipage}{\textwidth}\label{datanew}
  \centering
  \caption{Journal of the unpublished {\it HST} WFPC2 observations and QSOs measured parameters. \textsc{Notes} -- (8, 9) Nucleus and host galaxy apparent magnitudes in the observation filter. -- (10) Host galaxy elliptical radius. -- (11) Host galaxy apparent magnitude in the $R_{\rmn{Cousin}}$ band.}
\begin{tabular}{@{}ccccccccccc@{}}
\hline
\hline 
Name        &$z$&Proposal ID&Obs. Date&Filter&$T_{\rmn{exp}}$&Camera&$m_{\rmn{nuc}}$&$m_{\rmn{host}}$&$r_{\rmn{e}}$  &$R$  \\
(1)         & (2)        & (3)     & (4)      & (5)     & (6)   & (7) &(8)&(9)&(10)&(11)\\
\hline
%0637--752   &Elliptical  &15.94    &19.30     &15.90	&0.90'' &19.71\\
0850+440    &0.513&5949&07 Feb. '96&F702W&2400&WF3&16.93    &18.80     &1.03'' &19.21\\
1136--135   &0.554&6619&10 Jun. '97&F702W&2100&WF3&16.52    &18.82     &1.02'' &19.23\\
%1305+069    &Elliptical  &17.66    &20.36     &17.58    &1.13'' &19.84\\
\hline
\end{tabular}
\label{lastpage}
\end{minipage}
\end{table*}

\begin{figure*}
\centering
\begin{minipage}{\textwidth}
  \centering
\includegraphics[width=0.9\textwidth]{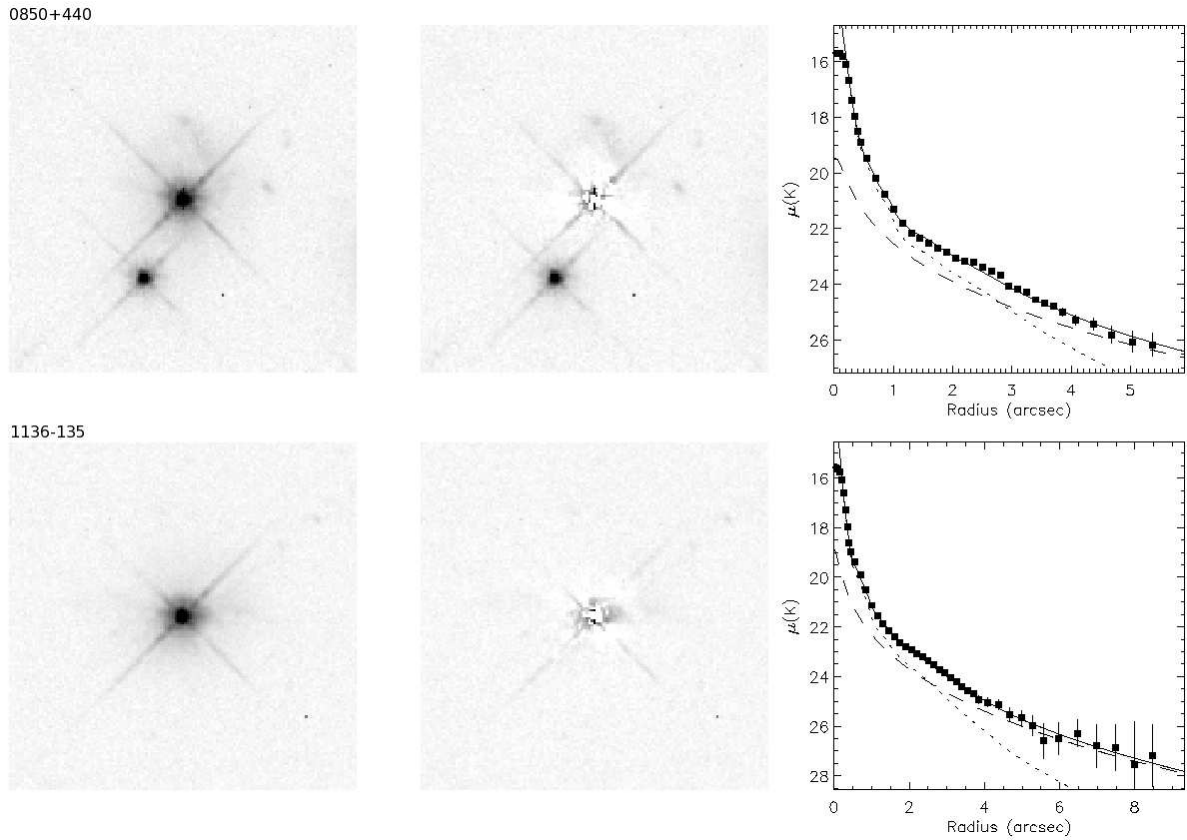}
\caption[]{QSOs images, 2-D residuals after the PSF subtraction, and radial profiles: data points (squares), PSF (dotted lines), galaxy model (dashed lines), best--fitting convolution functions (solid lines).}
\label{rp}
\end{minipage}
\end{figure*}
In Table 5 
we report the  journal of observations for 
the two objects without available  
measurement of the host galaxy from the literature. 
A two--dimensional analysis has been carried out using \textsc{AIDA}
(Astronomical Image Decomposition and Analysis, Uslenghi \& Falomo 2006, in preparation; see also Kotilainen et al.~2005), 
a software package specifically designed to
perform two dimensional model fitting of QSO images, providing
simultaneous decomposition into the nuclear and host galaxy components.

PSF modelling is by far the most critical step of analysing QSO host
galaxies. For HST images, the software package TinyTim (Krist \& Hook 1997)
can produce a PSF model that is very accurate in the inner part (up
to 1--2'') but it  does not
properly model the external fainter halo produced by the scattered
light (e.g. Scarpa et al.~2000). To account for this we 
therefore modelled the external part by adding an 
exponential term to the TinyTim PSF. 
The parameters of these components have been
evaluated by fitting the PSF model to a number of stars in the field. 

For each QSO we have then applied a mask to
exclude possible contamination from spurious sources, bad pixels, companions 
and other 
defects affecting the image. The background was estimated from the signal 
computed in a circular annulus centered on the object   and its
uncertainty obtained from the comparison  of the
values computed in annuli at different radii. 
The masked QSO images
were finally fitted with a model including both a point source 
and the host galaxy assuming a de Vaucouleurs law convolved with the PSF. 
The results of the best fit parameters are reported in Table 5.  
A fit with a disc galaxy yielded significant poorer results, consistently with the high luminosity of the nucleus (see Dunlop et al.~2003). 
In Fig.~\ref{rp} we show the QSOs images and residuals. The radial profiles of the images, of the PSF, of the galaxy and of the best--fitting models for the 2 objects are also shown.

\end{document}